\documentclass[conference]{IEEEtran}
\IEEEoverridecommandlockouts

\usepackage{cite}
\usepackage{mwe}
\usepackage{subfig}
\usepackage{setspace}
\usepackage{algorithm}
\usepackage{algorithmic}
\usepackage{adjustbox}
\usepackage{physics}

\usepackage{float}%
\usepackage{commath}
\usepackage{ifpdf}
\usepackage{cuted}
\usepackage{lipsum}
\usepackage{diagbox}
\usepackage{multirow}
\usepackage{romannum}
\usepackage{etoolbox}
\usepackage{mathtools}
\usepackage{amsfonts, amsmath, amsthm, amssymb}
\usepackage{bbm}
\usepackage{epsfig, graphicx, wrapfig, color}
\usepackage{hyperref}
\usepackage[normalem]{ulem}
\usepackage{xcolor}
\usepackage{mathtools}
\mathtoolsset{showonlyrefs}

\usepackage{mathtools}
\mathtoolsset{showonlyrefs}

\usepackage{annotate-equations}
\usepackage[skins,listings]{tcolorbox}

\usepackage{tikz}
\usetikzlibrary{tikzmark} 
\usepackage{hf-tikz} 

\usepackage{comment}

\allowdisplaybreaks

\newtheorem{theorem}{Theorem}

\newtheorem{definition}[theorem]{Definition}

\usepackage{xcolor}

\definecolor{2stageup}{RGB}{97,108,140} 
\definecolor{2stagedown}{RGB}{178,213, 155}

\begin{document}

\bstctlcite{IEEEexample:BSTcontrol}


\title{Higher-order Interpretations of Deepcode, a Learned Feedback Code}

\author{\IEEEauthorblockN{Y. Zhou$^1$, N. Devroye$^1$, Gy. Tur\'an$^{1,2}$,  and M. \v Zefran$^1$}
\IEEEauthorblockA{$^1$University of Illinois Chicago, Chicago, IL, USA}
\IEEEauthorblockA{$^2$HUN-REN-SZTE Research Group on AI}
\IEEEauthorblockA{\{yzhou238, devroye,  gyt, mzefran\}@uic.edu}
}


\maketitle

\begin{abstract}
We present an interpretation of Deepcode, a learned feedback code that showcases higher-order error correction relative to an earlier interpretable model~\cite{US-2024-ISIT-IEEE}. By interpretation, we mean succinct analytical encoder and decoder expressions (albeit with learned parameters) in which the role of feedback in achieving error correction is easy to understand. By higher-order, we mean that longer sequences of large noise values are acted upon by the encoder (which has access to these through the feedback) and used in error correction at the decoder in a two-stage decoding process. \footnote{This work was supported by NSF under awards 1900911, 2217023, and 2240532, and by the AI National Laboratory Program (RRF-2.3.1-21-2022-00004). The contents of this article are solely the responsibility of the authors and do not necessarily represent the official views of the NSF.}
\end{abstract}


\section{Introduction} \label{sec:introduction}
Shannon introduced the classical additive white Gaussian
noise (AWGN) channel with feedback in 1956~\cite{shannon1956zero} and proved a counterintuitive result: feedback does not increase the capacity of memoryless channels. Nevertheless, feedback can improve transmission reliability. A well-known example of an analytically constructed passive feedback coding method is the Schalkwijk and Kailath (SK) scheme~\cite{schalkwijk1966coding}. Designed for noiseless feedback, it is linear and achieves capacity with a doubly exponential error decay in block length. However, SK scheme fails when the feedback is noisy and suffers from numerical issues as the block length increases \cite{kim2020deepcode}. Accordingly, Chance and Love proposed a linear scheme known as the CL scheme for noisy feedback \cite{chance2011concatenated}. However, this scheme has been experimentally demonstrated to be suboptimal \cite{mishra2023linear} and one issue appears to be the restriction to linear codes. It has been proven that linear codes with noisy passive feedback fail to achieve a positive rate \cite{kim2007gaussian}. 



%

Recently, deep learning methods have been proposed to learn nonlinear feedback error correcting codes. Unlike traditional analytically constructed codes, these  deep-learned error-correcting feedback codes (DL-ECFCs)   are trained on large amounts of data to ``learn'' the encoders, decoders, or both by minimizing loss functions. These DL-ECFCs~\cite{kim2018deepcode, safavi2021deep, mashhadi2021drf, shao2023attentioncode, ozfatura2022all, kim2023robust} have demonstrated superior performance, particularly in channels with noisy feedback. However, they are considered ``black-box'' systems,
as their parameters,
without further interpretation, offer little insight into how the codes carry out error correction using the feedback.
We aim to understand and interpret the first code in the family of DL-ECFCs: Deepcode, which relies on recurrent neural network (RNN) and gated recurrent unit (GRU) architectures~\cite{kim2018deepcode}. Understanding Deepcode would increase confidence in the system and possibly provide a pathway to designing new analytical codes. The original Deepcode uses an RNN with 50 hidden states, making it difficult to interpret. In our prior work~\cite{US-2024-ISIT-IEEE}, through model reduction and analysis of system dynamics, we developed an (enc 2, dec 2) single-stage interpretable model\footnote{``enc $x$'' means the encoder uses $x$-order error correction components, and ``dec $y$'' means the decoder takes $y-1$ future steps into account. We expand on this notation and nomenclature in Section \ref{sec: interpretation}} based on Deepcode with 5 hidden states which offers comparable bit error rate (BER) performance with significantly reduced complexity. Additionally, it provides insights into how feedback is used to correct errors during transmission. However, it exhibits degraded BER performance at low forward Signal-to-Noise Ratio (SNR) as it only utilizes feedback from the previous time step, and cannot handle longer noise events.

{\bf Contribution}: In this work, we generalize our previous model to a more powerful and interpretable model for both the encoder and decoder. The encoder incorporates feedback from multiple time steps. Based on this, we develop a two-stage decoder to effectively utilize the information embedded in the noisy received codewords. This new (enc 3, dec 4) two-stage interpretable model can handle longer noise events, thereby improving error correction capabilities. Additionally, we introduce an error analysis method to trace noise events, which helps identify limitations of existing methods and provides insights for further improvements.

{\bf Notation}: Subscripts $i$ are used to denote time indices. Any interval $[a:b]$ (where $a$ and $b$ are positive integers, with $a < b$) represents a sequence of integers $[a, a+1, \ldots, b]$. Vectors are in bold, with superscripts indicating their lengths. $\text{SNR}_f$ and $\text{SNR}_{fb}$ denote the Signal-to-Noise Ratios for the forward and feedback channels, respectively. $\mathbb{R}^n$ represents $n$-dimensional real vectors, while $\mathbb{F}_2$ denotes the finite field with elements $0$ and $1$. The function $\mathbb{I}(x) = 1$ if the argument $x$ is true (non-negative), and 0 otherwise. The symbol $++$ denotes excessively large positive, while $--$ denotes excessively large negative. $\uparrow$ means increase, while $\downarrow$ means decrease. {\abs{\cdot} denotes the absolute value.}

\section{Background}
This section provides a brief overview of the system model and the original Deepcode \cite{kim2018deepcode} that we will be interpreting.

\subsection{System model}
\begin{figure}[htbp]
  \centering
  \begin{tikzpicture}
    \node (encoder) at (2,0) [draw, rectangle] {Encoder $f_{\textcolor{red}{\theta}}$};
\node (plus1) at (4,0) [draw, circle] {$+$};
\node (decoder) at (6.3,0) [draw, rectangle] {Decoder $g_{\textcolor{red}{\phi}}$};
\node (plus2) at (4,-1.5) [draw, circle] {$+$};
\node (delay) at (5.1,-0.8) [draw, rectangle] {Delay};
\draw[->] (0,0) -- (encoder) node[midway, above] {$\mathbf{b}$};
\draw[->] (encoder) -- (plus1) node[midway, above] {$x_i$};
\draw[->] (plus1) -- (decoder) node[midway, above] {$y_i$};
\draw[->] (decoder) -- (8,0) node[midway, above] {$\mathbf{\hat{b}}$};
\draw[->] (4,0.7) -- (plus1) node[pos=0.2, above] {$n_i\sim \mathcal{N}(0,\sigma_f^2)$};
\draw (5.1,0) -- (delay);
\draw (delay) -- (5.1,-1.5) node[midway, right] {$y_{i-1}$};
\draw[->] (5.1,-1.5) -- (plus2);
\draw (plus2) -- (2,-1.5) node[midway, above] {$\tilde{y}_{i-1}$};
\draw[->] (2,-1.5) -- (encoder);
\draw[->] (4,-2.2) -- (plus2) node[pos=0.2, below] {$\tilde{n}_{i-1}\sim \mathcal{N}(0,\sigma_{fb}^2)$};
\draw[thick] (0.3,0.5) rectangle (3.5,-1.6);
\draw[thick] (4.5,0.5) rectangle (7.8,-1.6);
\node at (2,0.7) {Transmitter};
\node at (6.2,0.7) {Receiver};
  \end{tikzpicture}
  \caption{\small Learned parameters $\theta$ and $\phi$ in the encoding $f_\theta$ and decoding $g_\theta$ functions for transmission over the AWGN channel with passive noisy feedback. Forward noise is i.i.d. Gaussian ${\cal N}(0,\sigma_f^2)$ while feedback noise is also i.i.d. Gaussian ${\cal N}(0,\sigma_{fb}^2)$. }
  \vspace{-2mm}
  \label{fig:awgn}
\end{figure}
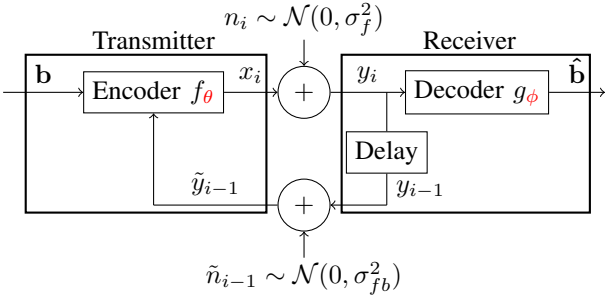

This study focuses on the AWGN channel with passive noisy feedback, as shown in Fig. \ref{fig:awgn}. The transmitter sends $K$ message bits $\mathbf{b} \in \mathbb{F}_2^K$ through $N$ channel uses, where the code rate is $r = \frac{K}{N}$. At time $i \in \{1, \ldots, N\}$, the receiver receives the noisy outputs $y_{i} = x_{i} + n_{i}$, where $x_{i} \in \mathbb{R}$ are the sent symbols (of the codewords ${\bf x}$) and $n_{i} \sim \mathcal{N}(0,\sigma_{f}^2)$ are independent and identically distributed (i.i.d.) Gaussian noises. We consider the average power constraint $\frac{1}{N}\mathbb{E}\left[\lVert \mathbf{\mathbf{x}} \rVert_2^2\right]\leq 1$. The receiver sends the received {symbols} to the transmitter through a noisy channel with one unit delay: $\tilde{y}_{i-1} = y_{i-1} + \tilde{n}_{i-1}$, where $\tilde{n}_{i-1} \sim \mathcal{N}(0,\sigma_{fb}^2)$.

For each transmission, the encoding function $f_\theta$\footnote{In general this may depend on $i$, hence $f_{\theta, i}: \mathbb{F}_2^K \times \mathbb{R}^{i-1} \rightarrow \mathbb{R}$ but we simplify the notation and simply refer to this as $f_\theta$, where the dependence on $i$ is understood.} 
maps the message bits and past noisy feedback to the {symbol} $x_i = f_{\theta}(\mathbf{b}, \mathbf{\tilde{y}}^{i-1})$. After $N$ transmissions, the decoder $g_\phi: \mathbb{R}^N \rightarrow \mathbb{F}_2^K$ estimates the message bits from the received {noisy codewords}, i.e. $\mathbf{\widehat{b}} = g_{\phi}(\mathbf{y}^N)\in \mathbb{F}_2^K$.  One metric for communication reliability is the $\text{BER}=\frac{1}{K}\sum_{i=1}^{K}\mathbb{P}(b_i\neq \widehat{b}_i)$. In general, the encoding and decoding functions can be anything. In Deepcode, they are neural networks that are jointly trained to minimize the binary cross-entropy (BCE), which acts as a surrogate metric for BER. See, for example, how BCE may be sandwiched by BER in~\cite[Proposition 3]{US-2023-ISIT}.

\subsection{Deepcode structure}

\begin{figure}[ht]
\vspace{-4mm}
    \centering
    \includegraphics[width=\columnwidth]{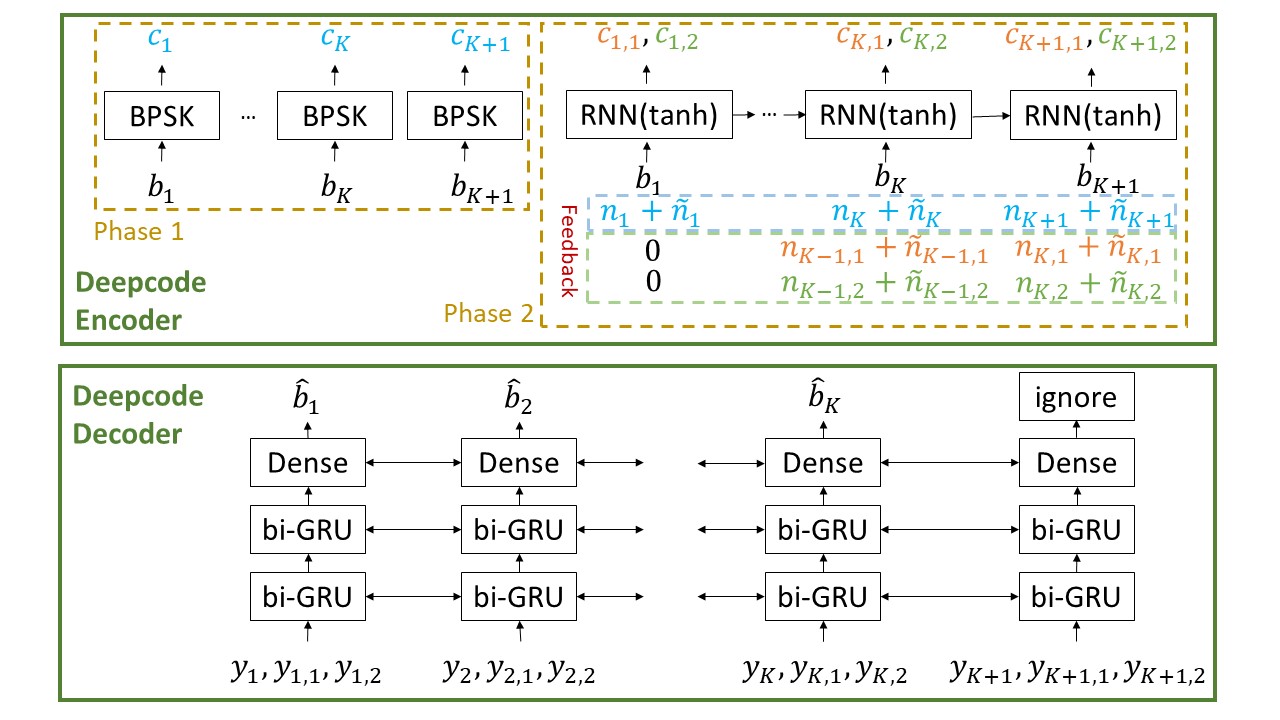}
    \caption{Deepcode encoder (above) and decoder (below). Here, $i \in \{1, \ldots, K + 1\}$ because of zero padding. We omit power allocation and normalization for simplicity.}
    \vspace{-4mm}
    \label{fig:deepcode_structure}
\end{figure}

Deepcode \cite{kim2018deepcode}, one of the first learned nonlinear feedback codes, uses an architectural structure and two-phase communication structure as shown in  Fig. \ref{fig:deepcode_structure}.

 Deepcode is a rate $r = \frac{1}{3}$ code which takes $K = 50$   input message bits $\mathbf{b}$ and produces a sequence of $N=150$ output {codewords $\mathbf{x}$} over two encoding  phases. The first phase is uncoded transmission of these 50 bits, simply applying  BPSK modulation $\mathbf{c} = 2\mathbf{b}-1 \in \{\pm 1\}^K$ to the message bits, for transmission over the AWGN channel. The noisy received {symbols} are directly sent back over the (noisy) feedback channel generating noisy feedback symbols  $\tilde{y}_1, \ldots, \tilde{y}_K$ at the transmitter, which are used in the second phase. The second phase uses an RNN with $\tanh$ activation, containing $N_h = 50$ hidden states, and a linear combination layer to sequentially generate $2K = 100$ {parity} symbols, $c_{i,1}$ and $c_{i,2}$, where $i \in \{1, \ldots, K\}$. For the $i$-th bit, the RNN accepts the message bit $b_i$, the feedback noise from the first phase $\tilde{y}_i - c_i$, and the delayed feedback noise from previous second phase sent symbols $\tilde{y}_{i-1,1} - c_{i-1,1}$ and $\tilde{y}_{i-1,2} - c_{i-1,2}$ and then spits out two coded parity symbols $c_{i,1}$ and $c_{i,2}$ for transmission over the AWGN channel. Based on the received noisy channel outputs $\mathbf{y}^N = \left[y_1,\cdots,y_K, y_{1,1}, y_{1,2},y_{2,1}, y_{2,2},\cdots, y_{K,1}, y_{K,2} \right]$, the decoder uses a two-layered bidirectional GRU to estimate the message bits $\mathbf{\hat{b}}$. It does this by forming a soft output (probability that bit $i$ is 1) and using the sigmoid function (a threshold of 0.5) to produce the estimate. If the soft output is greater than the threshold, the estimated bit is 1; otherwise, it is 0.

Deepcode applies zero padding to reduce the error of the last bit and employs power allocation to balance the error, followed by a normalization layer to ensure the codewords satisfy the power constraint. To simplify our interpretation, we  first focus on the noiseless feedback case ($\tilde{n}_i = \tilde{n}_{i-1,1} = \tilde{n}_{i-1,2} = 0$) and then generalize to noisy feedback case.


\section{Higher-order interpretable model}\label{sec: interpretation}
In our prior work~\cite{US-2024-ISIT-IEEE} we presented the (enc 2, dec 2) single-stage interpretable model based on Deepcode with $N_h = 5$ hidden states, i.e. we re-trained Deepcode using 5 hidden states in both the encoder RNN and decoder GRU rather than 50, significantly reducing the number of learned parameters. We then obtained an \textit{interpretable} model by  finding a simpler and more understandable nonlinear expression (through influence, pruning and outlier analysis) for both the encoder and the decoder that yields comparable BER performance to the original learned Deepcode.



{We build on the approach of \cite{US-2024-ISIT-IEEE}, which introduced outlier analysis to understand how the encoder performs error correction using feedback. Additionally, we identify the noise pattern leading to decoding errors by conducting a correlation analysis to find relevant features, followed by K-means clustering to classify them.}




\begin{definition}

    An (enc $x$, dec $y$) interpretable model is defined as follows, for $x\in \{1,2,3\}$:
    \begin{itemize}
        \item The encoder applies $x$-order error correction, which follows a hierarchical error-correction method:
        \begin{enumerate}
            \item $x=1$ or first-order correction addresses large errors in $n_i$;
            \item $x=2$ or second-order correction extends to correcting large errors in $n_i$ along with combinations of large ranges of $n_{i,1}$, and $n_{i,2}$;
            \item $x=3$ or third-order correction extends to correcting large errors in $n_i$, the same specific combinations of large errors in $n_{i,1}$, $n_{i,2}$ and finally specific large ranges of $n_{i+1,1}$, and $n_{i+1,2}$.
        \end{enumerate}
        {Notice that higher-order error correction addresses noise events that are subsets of those addressed by lower-order error correction.}
        \item The decoder uses $y-1$ time-step parity bit sums ($c_{i+y-1,1} + c_{i+y-1,2}$) for decoding $b_i$. ``Two-stage decoding'' involves both forward and backward passes, while ``single-stage decoding'' does not.
    \end{itemize}

\end{definition}

{This definition can be extended beyond $x\in \{1,2,3\}$ but is left for future work.  {We envision higher orders would be able to handle longer and longer sequences of large noise events.} 
Here we extend our previous interpretations to a re-trained \textbf{(enc 3, dec 4) two-stage interpretable model} based on Deepcode with $N_h = 7$ hidden states, which offers better performance while still remaining more manageable to interpret than the original Deepcode with $N_h=50$ hidden states. {Compared to enc 2 (considered in \cite{US-2024-ISIT-IEEE} but not with that nomenclature), enc 3 enables more refined error correction.
}

\subsection{Encoder Interpretation}\label{sec:encoder inter}
We first discussed the enc 3 in Appendix D of ~\cite{zhou2024interpreting}, but did not consider the decoder. Here, we provide a more detailed analysis of both the encoder and decoder. In the following, we assume without loss of generality that the learned parameters (in \textcolor{red}{red}) are positive, except for the bias. If these learned parameters are negative we can create equivalent codes by corresponding sign changes, see Appendix F of ~\cite{zhou2024interpreting}.

In summary, our enc 3 interpretable analytical encoder looks as follows. For each $i\in \{1,2,\cdots,50\}$, the coded symbols output by the encoding function $f_\theta$ are: 
\begin{align}
\text{Phase 1: } \;\; c_i  = & 2b_{i} -1  \\
\label{parity 1}
\text{Phase 2: } \;\;  c_{i, 1} = & \eqnmarkbox[purple]{parity1st}{\textcolor{red}{e_1}n_{i}\mathbb{I}\left(-(2b_i-1)n_{i}\right)} \eqnmarkbox[cyan]{parity2st}{- \textcolor{red}{e_2}h_{i,4} - \textcolor{red}{e_2} h_{i,5}} \nonumber \\
& \eqnmarkbox[orange]{parity3st}{- \textcolor{red}{e_3}h_{i,6} + \textcolor{red}{e_3}h_{i,7}} \\[2.5mm]
\label{parity 2}
c_{i, 2} = & \eqnmarkbox[purple]{paritys1st}{- \textcolor{red}{e_1}n_{i}\mathbb{I}\left(-(2b_i-1)n_{i}\right)} \eqnmarkbox[cyan]{paritys2st}{- \textcolor{red}{e_2}h_{i,4} - \textcolor{red}{e_2} h_{i,5}} \nonumber\\
& \eqnmarkbox[orange]{paritys3st}{- \textcolor{red}{e_3}h_{i,6} + \textcolor{red}{e_3}h_{i,7}} 
\end{align}
where $e_1$, $e_2$ and $e_3$ are learned coefficients. {We next show how this may enable what we term as three ``orders'' of error correction. Lower orders correct shorter strings of noise events; higher orders enable longer sequences of noise events  to be corrected, enabling lower BERs.} 
\annotate[yshift=0.5em, xshift = 0em]{right}{parity1st}{first-order}
\annotate[yshift=0.5em, xshift = 0em]{right}{parity2st}{second-order}
\annotate[yshift=0.2em, xshift = 0.6em]{below,}{parity3st}{third-order}

\subsubsection{First-order Error Correction}
When forming the parity bits, {the linear combination of the first three hidden states in the encoder RNN ($\textcolor{red}{a_1}h_{i,1} + \textcolor{red}{a_2}h_{i,2} + \textcolor{red}{a_3}h_{i,3}$) may be approximated by the first-order component which sends a scaled version of phase 1 noise, if needed.} If $b_i=0$ and phase 1 noise is negative (or $b_i=1$ and phase 1 noise is positive), binary detection would accurately estimate the bit without requiring additional information to be transmitted. Otherwise, subtracting the two parity bits $c_{i,1} - c_{i,2}$ isolates the ``\textcolor{purple}{purple}'' part, which contains the scaled version of the phase 1 noises for first-order error correction {i.e. if the receiver could do this, it could cancel out $n_i$ experienced by the uncoded bit and correctly decode it. However, during transmission, the receiver only has access to  a noisy version of the subtraction result, i.e. can only create $y_{i,1} - y_{i,2} = c_{i,1} - c_{i,2} + n_{i,1} - n_{i,2}$. Thus, the receiver's estimate of $n_i$ may be corrupted by phase 2 noises. If these are small, first order error correction is achieved, but if $n_i, n_{i,1}, n_{i,2}$ are large, this would  require additional error correction to address the phase 2 noises.}



\subsubsection{Second-order Error Correction}
The ``\textcolor{cyan}{cyan}'' part $h_{i,4}$ and $h_{i,5}$ (which are RNN hidden states that we are interpreting) enable second-order error correction, are defined as: 
\begin{align}
 \text{If } b_{i-1}&=0, \text{ then } h_{i,5}=-1 \text{ and }\\
 h_{i,4} &= \tanh{\left(-\textcolor{red}{k_1}n_{i-1} + \textcolor{red}{k_2}n_{i-1, 1} - \textcolor{red}{k_3}n_{i-1, 2} + \textcolor{red}{k_4}\right)}  \label{eq:h4}\\
 \text{If } b_{i-1}&=1, \text{ then }  h_{i,4} =  1 \text{ and }\\
 h_{i,5} &=  \tanh{\left(-\textcolor{red}{k_1}n_{i-1} + \textcolor{red}{k_2}n_{i-1, 1} \label{eq:h5}
 - \textcolor{red}{k_3}n_{i-1, 2}- \textcolor{red}{k_4}\right)} 
 \end{align}
where $k_1$, $k_2$, $k_3$, and $k_4$ are learned coefficients. As discussed in~\cite{US-2024-ISIT-IEEE}, $h_{i+1,4}$ and $h_{i+1,5}$ eliminate phase 2 noises at time $i$, or enable second-order error correction, by summing the one-time-step parity bits $c_{i+1,1} + c_{i+1,2}$, which cancels out the first-order term. In most cases, $h_{i+1,4}$ are mostly 1 and $h_{i+1,5}$ are mostly -1. Their values cancel out when forming the parity bits. However, if phase 2 noises ($n_{i,1} - n_{i,2}$) are large enough to perturb the estimate of $n_i$ and affect the decoding result, either $h_{i+1,4}$ or $h_{i+1,5}$ will exhibit outliers to aid in error correction. Briefly, $h_{i+1,4}$ corrects noise when the message bit $b_i$ is 0 by increasing the one-time-step parity bits, while $h_{i+1,5}$ targets the message bit 1 by decreasing the one-time-step parity bits, as shown in Table \ref{encoder:paritychange}. 

Based on these two error correction orders, the corresponding dec 2 single-stage decoding (where 2 indicates considering one-time-step parity bit sum)~\cite{US-2024-ISIT-IEEE} is formulated as follows:
\begin{align}
\label{eq:dec2}
o_{i,j}  = &\tanh\left(\textcolor{red}{d_{j,1}}y_i - \textcolor{red}{d_{j,2}}\eqnmarkbox[purple]{node4}{(y_{i, 1}- y_{i, 2})} \right. \nonumber \\ 
& \quad \left.- \textcolor{red}{d_{j,3}}\eqnmarkbox[cyan]{node5}{(y_{i+1, 1} + y_{i+1, 2})} +  \textcolor{red}{d_{j,4}}\right)  \\ 
\hat{b}_{i}  = & \begin{cases}0 & \text{if } D_i < 0.5, \\ 1 & \text{if } D_i \geq 0.5. \end{cases} \quad D_i  = \sigma\left(\sum_{j=1}^{N_l=5}\textcolor{red}{l_j}o_{i,j}\right)
\end{align}
Here, $\sigma$ represents the sigmoid function. $o_{i,j}$ estimates the input bit by removing phase 1 and phase 2 noises from the uncoded received $y_i$ using different combinations (subtracting or summing) of noisy parity bits. We generate the soft output $D_i$ by applying the sigmoid to a linear combination of $o_{i,j}$ with $N_l = 5$, then threshold $D_i$ to obtain the estimated bit. If the message bit $b_i$ is 0, second-order error correction (outlier in $h_{i+1,4}$) increases the one-time-step (noisy) parity sum $(y_{i+1, 1} + y_{i+1, 2})$, leading to a negative $o_{i,j}$ and a smaller probability $D_i$. Conversely, for message bit 1, an outlier in $h_{i+1,5}$ causes $o_{i,j}$ to become positive for correct decoding.





\subsubsection{Third-order Error Correction}
Up until now, we have  re-capped the \textbf{(enc 2, dec 2) single-stage interpretable model} with $N_h=5$ hidden states. Although it is simple and achieves good BER performance, it is still an order of magnitude worse than Deepcode with 7 hidden states at low forward SNR, as shown in Fig. \ref{fig:7berperformance}. We now interpret this more sophisticated encoder and decoder, { which has a similar structure to the $N_h=5$ one, but intuitively, with the ability to conduct higher-order error correction (i.e. longer sequences of noise events)}

\textit{Interpretable tools to understand the noise events:} We collect error instances where the (enc 2, dec 2) single-stage interpretable model decodes incorrectly, but Deepcode with 7 hidden states decodes correctly using the same input message bits and noises (at forward $\text{SNR}_f  = 0\text{dB}$ and noiseless feedback). Specifically, we gather data around the error location $i$ (excluding the boundary bits), including message bits $\mathbf{b}_{i-4:i+4}$, phase 1 noises $\mathbf{n}_{i-4:i+4}$, and phase 2 noises $\mathbf{n}_{i-4:i+4,1}$, $\mathbf{n}_{i-4:i+4,2}$. 
We then compute the pairwise correlation of true message bits $b_i$ with other features. The five features with the largest absolute correlation coefficients are $\rho_{b_{i}, n_{i+1,1}} = 0.956$, $\rho_{b_{i}, n_{i+1,2}} = 0.952$, $\rho_{b_{i}, n_{i,1}} = 0.849$ and $\rho_{b_{i}, n_{i,2}} = -0.840$, $\rho_{b_{i}, n_{i}} = -0.799$. Other features have absolute values less than $0.1$, indicating they are less relevant.


{To identify the error patterns, we applied K-means clustering to the relevant features ($n_{i}$, $n_{i,1}$, $n_{i,2}$, $n_{i+1,1}$, and $n_{i+1,2}$) that cause errors in the (enc 2, dec 2) single-stage interpretable model but are corrected by the 7 hidden states Deepcode.} After testing different numbers of clusters, we found that dividing the data into two groups best describes its distribution ({one for $b_i=0$ and one for $b_i=1$}). From Fig. \ref{fig:errorcluster}, we observed that the features leading to errors follow a pattern: one cluster exhibits $n_{i} ++ $, $n_{i,1} --$, $n_{i,2} ++$ (which may cause an outlier in $h_{i+1,4}$ from Table. \ref{encoder:paritychange}), while $n_{i+1,1}$ and $n_{i+1,2}$ are excessively negative. The other cluster shows the opposite pattern: $n_{i} -- $, $n_{i,1} ++$, $n_{i,2} --$ (resulting in an outlier of $h_{i+1,5}$), with $n_{i+1,1}$ and $n_{i+1,2}$ being excessively positive.

{The clustering results indicate that decoding issues arise from large same-sign phase 2 noises at time $i+1$. At the receiver, we only have access to the noisy one-time-step parity bit sum: $y_{i+1, 1} + y_{i+1, 2} = c_{i+1, 1} + c_{i+1, 2} + n_{i+1, 1} + n_{i+1, 2}$. For message bit $b_i = 0$, we expect $o_{i,j}$ in \eqref{eq:dec2} to be as negative as possible for correct decoding. If phase 2 noises at time $i+1$ have the same sign and are significantly negative ($c_{i+1, 1}^{\uparrow} + c_{i+1, 2}^{\uparrow} + n_{i+1, 1}^{--} + n_{i+1, 2}^{--}$), these noises prevent the increase of the parity bit sum, pushing $o_{i,j}$ towards the positive direction, thus blocking error correction and potentially causing decoding errors. A similar case occurs for message bit 1 when the phase 2 noises are significantly positive ($c_{i+1, 1}^{\downarrow} + c_{i+1, 2}^{\downarrow} + n_{i+1, 1}^{++} + n_{i+1, 2}^{++}$), blocking the decrease of the parity bit sum necessary for error correction.}



\begin{figure}[ht]
\vspace{-4mm}
    \centering
    \includegraphics[width=\columnwidth]{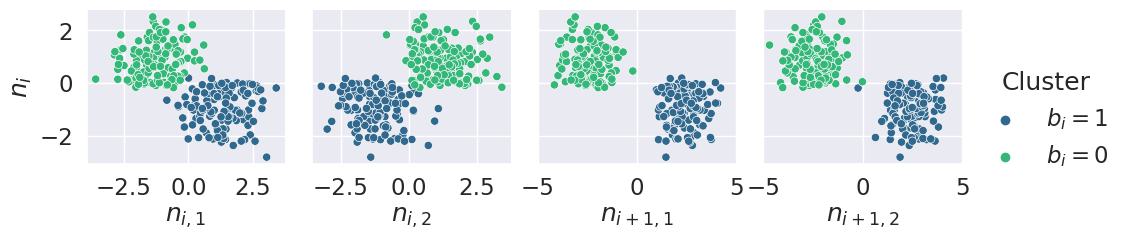}
    \caption{Clustering error features leading to decoding errors in the (enc 2, dec 2) single-stage interpretable model corrected by Deepcode with $7$ hidden states.}
    \vspace{-4mm}
    \label{fig:errorcluster}
\end{figure}


 \begin{figure*}[t] 
\vspace{-4mm}
    \centering
    \includegraphics[width=\textwidth]{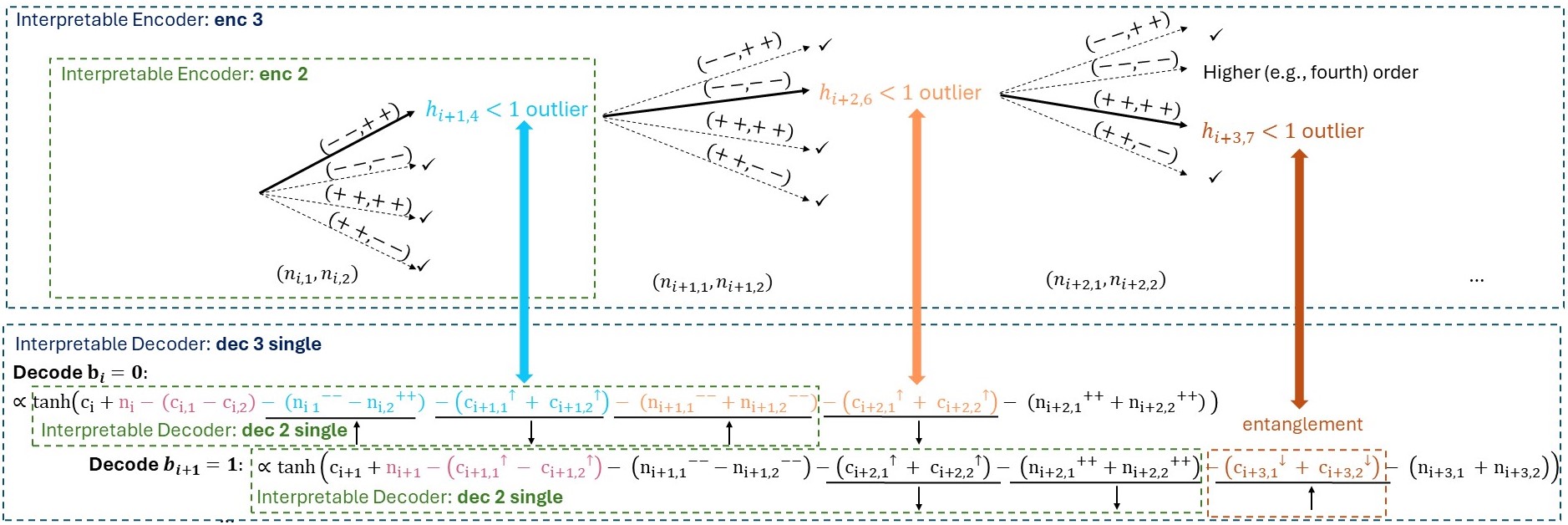}
    \caption{An example of extended noise events occurring for $b_i=0$. The solid line represents noise events that will cause outliers and alter the future parity bits for error correction, whereas the dashed line indicates that no further error correction is needed. We omit the coefficients in the decoding for simplicity.}
    \vspace{-4mm}
    \label{fig:encodertrace}
\end{figure*}
Therefore, we introduce the new ``\textcolor{orange}{orange}'' component, which represents the third-order error correction. {$h_{i+2,6}$ and $h_{i+2,7}$ improve performance by addressing large same-sign phase 2 noises in $n_{i+1,1}$ and $n_{i+1,2}$}. Additionally, they form entanglements, becoming mutually dependent on each other's previous values, which prevents over-correction for subsequent bits. Our interpretable approximations of $h_{i,6}$ and $h_{i,7}$ are defined as follows:
\begin{align}
h_{i,6} & = \tanh\left(\textcolor{red}{m_1}n_{i-1, 1} + \textcolor{red}{m_2}n_{i-1, 2} + \textcolor{red}{m_3}h_{i-1,4} \right. \nonumber \\
& \quad \left. + \textcolor{red}{m_4}h_{i-1,7} + \textcolor{red}{m_5}\right) \\
h_{i,7} & = \tanh\left(-\textcolor{red}{m_1}n_{i-1, 1} - \textcolor{red}{m_2}n_{i-1, 2} - \textcolor{red}{m_3}h_{i-1,5} \right. \nonumber \\
& \quad \left. + \textcolor{red}{m_4}h_{i-1,6} + \textcolor{red}{m_5}\right)
\end{align}
where $m_1$ to $m_5$ are learned parameters. 

Without {``entanglement'' (which we will use later as a comparison point)}, the hidden states are defined as:
\begin{align}
h_{i,6}^{'} & = \tanh\left(\textcolor{red}{m_1}n_{i-1, 1} + \textcolor{red}{m_2}n_{i-1, 2} + \textcolor{red}{m_3}h_{i-1,4}   + \textcolor{red}{m_5}\right) \\
h_{i,7}^{'} & = \tanh\left(-\textcolor{red}{m_1}n_{i-1, 1} - \textcolor{red}{m_2}n_{i-1, 2} - \textcolor{red}{m_3}h_{i-1,5} +\textcolor{red}{m_5}\right) 
\end{align}

The values for $h_{i,6}$ ($h_{i,6}^{'}$) and $h_{i,7}$ ($h_{i,7}^{'}$) are predominantly 1. If the noise levels are small enough not to affect decoding results, there will be no outliers, and the values will offset in the parity bits. When they have outliers, they affect the parity bits as shown in the Table \ref{encoder:paritychange}. We sum the two-time-step parity bits $c_{i+2, 1} + c_{i+2, 2}$ to mitigate same-sign phase 2 noises at time $i+1$. This adjustment is expected to counteract the effect of same-sign phase 2 noises and steer $o_{i,j}$ towards negative values for message bit 0 or positive values for message bit 1, respectively.

\begin{table}[htbp]
\centering
\begin{adjustbox}{max width = \columnwidth}
\begin{tabular}{|c|c|c|}
\hline
$c_{i,j}$ & outlier & noise combination \\ \hline
\multirow{3}{*}{$\big\uparrow$} & $h_{i,4} <1 (\downarrow)$ & $b_{i-1} = 0, n_{i-1} ++, n_{i-1,1} --, n_{i-1,2} ++$ \\ \cline{2-3} 
 & \multirow{2}{*}{$h_{i,6} < 1 (\downarrow)$} & $h_{i-1,4} < 1 (\downarrow), n_{i-1,1} --, n_{i-1,2} --$ \\ \cline{3-3} 
 &  & ${h_{i-1,7} < 1 (\downarrow)}, n_{i-1,1} --, n_{i-1,2} --$ \\ \hline
\multirow{3}{*}{$\big\downarrow$} & $h_{i,5} > -1 (\uparrow)$ & $b_{i-1} = 1, n_{i-1} --, n_{i-1,1} ++, n_{i-1,2} --$ \\ \cline{2-3} 
 & \multirow{2}{*}{$h_{i,7} < 1(\downarrow)$} & $h_{i-1,5} > -1(\uparrow), n_{i-1,1} ++, n_{i-1,2} ++$ \\ \cline{3-3} 
 &  & $h_{i-1,6} < 1 (\downarrow), n_{i-1,1} ++,n_{i-1,2}++ $ \\ \hline
\end{tabular}
\end{adjustbox}
\caption{\label{encoder:paritychange}The effect of outliers on parity bits $c_{i,j}$, where $j\in\{1, 2\}$. Meeting two or three conditions may lead to an outlier, even if the noise condition is not fully satisfied, as long as the trend is followed.}
\vspace{-2mm}
\end{table}

{In conclusion, the first-order correction removes phase 1 noise $n_i$ from the uncoded signal $y_i$ when $b_i=0, n_i++$ (or $b_i=1, n_i--$). However, the estimate of $n_i$ is corrupted by phase 2 noise at the receiver. When $b_i=0, n_i++, n_{i,1}--, n_{i,2}++$ (or $b_i=1, n_i--, n_{i,1}++, n_{i,2}--$), the second-order correction adjusts the one-time-step parity bit sum $c_{i+1,1}+ c_{i+1,2}$ to eliminate phase 2 noise at time $i$. For longer noise events, if phase 2 noise at time $i+1$ prevents the adjustment of the one-time-step parity bit sum, the third-order correction adjusts the two-time-step parity bit sum for additional correction.}

We conducted an ablation study on the encoder components. Table \ref{table:encoder compare} presents the BER performance using the Deepcode decoder with 5 hidden states across various encoder configurations. Interpretable enc 3 without entanglement improves BER by half-order-of-magnitude compared to interpretable enc 2, while adding entanglement enhances performance further by building longer dependencies between codewords.


\begin{table}[ht]
\centering
\begin{adjustbox}{max width = \columnwidth}
\begin{tabular}{|l|l|l|}
\hline
Encoder & Decoder & BER  $\text{SNR}_f$ $0\text{dB}$    \\ \hline
deep (5) & deep (5) & $6.375e - 05$ \\ \hline
interpret. enc 2  & deep (5)      & $7.616e-05$     \\ \hline
interpret. enc 3 w/o eg.  & deep (5) & $1.367e-05$ \\ \hline
interpret. enc 3 & deep (5) & $6.416e-06$ \\ \hline
deep (7) & deep (5) & $5.419e-06$ \\ \hline
\end{tabular}
\end{adjustbox}
\caption{\label{table:encoder compare}BER performance of different encoders (noiseless feedback). The number of hidden states is indicated in parentheses. ``interpret. enc 3 w/o eg.'' refers to enc 3 without entanglement.}
\vspace{-3mm}
\end{table}

\subsection{Decoder Interpretation}\label{decoder}

Given this, an intuitive decoding approach for enc 3 is to extend single-stage dec 2 in \eqref{eq:dec2} by one more time step, which we refer to as ``single-stage dec 3''. Additionally, instead of using $N_l = 5$, we will use $N_l = 7$. 
\begin{align}
    o_{i,j}  = &\tanh\left(\textcolor{red}{d_{j,1}}y_i - \textcolor{red}{d_{j,2}}{(y_{i, 1}- y_{i, 2})} - \textcolor{red}{d_{j,3}}{(y_{i+1, 1} + y_{i+1, 2})}  \right. \nonumber \\
& \quad \left. - \textcolor{red}{d_{j,4}}\eqnmarkbox[orange]{decmore}{(y_{i+2, 1} + y_{i+2, 2})} +  \textcolor{red}{d_{j,5}}\right) \label{eq:7naivedecoder}
\end{align}


{Fig. \ref{fig:encodertrace} illustrates three orders of error correction, demonstrating how parity bits adjust under different noise scenarios for $b_i = 0$. To correctly decode, $o_{i,j}$ for message bit $b_i=0$ should be negative. When decoding $b_i$, we initially estimate phase 1 noise by subtracting the current parity bit \textcolor{purple}{$c_{i,1}-c_{i,2}$}. However, if current parity bits carry phase 2 noises like $n_{i,1}--, n_{i,2}++$, this increases $o_{i,j}$, corrected by second-order adjustment \textcolor{cyan}{$c_{i+1,1}+c_{i+1,2}$}. Excessively negative phase 2 noises at $i+1$ can undermine this, necessitating third-order correction, which increases \textcolor{orange}{$c_{i+2,1}+c_{i+2,2}$}. If phase 2 noise at $i+2$ continues to be extremely negative (in rare cases where we have long noise event sequences), fourth-order or higher correction is required to achieve an even lower BER. }

{\textbf{Entanglement}:  In the (enc 2, dec 2) single-stage interpretable model, decoding only involves two orders of error correction: ``\textcolor{purple}{first-order}'' and ``\textcolor{cyan}{second-order}''. Separating these two orders is straightforward, achieved through subtraction or summing. However, in the (enc 3, dec 3) single-stage interpretable model, decoding requires incorporating information from three orders. At the receiver, it is difficult to distinguish whether the change in the sum of parity bits arises from ``\textcolor{cyan}{second-order}'' or ``\textcolor{orange}{third-order}'' corrections. In Fig. \ref{fig:encodertrace}, the third-order for $b_i$ $(c_{i+2,1} + c_{i+2,2})$ increases to correctly decode $b_i$. However, if $b_{i+1}$ is message bit 1, the increase in $(c_{i+2,1} + c_{i+2,2})$, which is also considered as second-order correction for $b_{i+1}$, can push $o_{i+1,j}$ towards the negative direction, potentially leading to an decoding error. This situation worsens with significantly positive phase 2 noises at time $i+2$, prompting entanglement (outliers in $h_{i+3, 7}$) to counteract over-correction. Thus, the enc 3 with entanglement performs better.}

Although dec 3 single-stage decoder in \eqref{eq:7naivedecoder} improves the BER performance by half an order of magnitude compared to dec 2 (see Table \ref{table:decoder compare}), it is not sufficient to fully exploit the information contained in the enc 3 encoder. The longer dependency between codewords makes decoding much more difficult because changes in the parity bits involve more cases. When comparing instances where single-stage dec 3 decodes incorrectly but the deep learned decoder with 5 hidden states decodes correctly using the same encoder (interpretable enc 3), we identified the following issue: because the decoding window spans three time steps, an error in one bit causes over-correction in the surrounding bits. There are mainly four types of errors in Table \ref{tab:4type} that need to be considered.
\begin{table}[htbp]
\centering
\begin{adjustbox}{max width = \columnwidth}
\begin{tabular}{|c|c|ccc|cc|c|c|}
\hline
Type & $b_i$ & \multicolumn{3}{c|}{noises $i$} & \multicolumn{2}{c|}{noises $i+1$} & $c_{i+1,j}$ & $c_{i+2,j}$ \\ \hline
 &  & \multicolumn{1}{c|}{$n_i$} & \multicolumn{1}{c|}{$n_{i,1}$} & $n_{i,2}$ & \multicolumn{1}{c|}{$n_{i+1,1}$} & $n_{i+1,2}$ &  &  \\ \hline
1 & 0 & \multicolumn{1}{c|}{$++$} & \multicolumn{1}{c|}{$--$} & $++$ & \multicolumn{1}{c|}{} &  & $\uparrow$ &  \\ \hline
2 & 1 & \multicolumn{1}{c|}{$--$} & \multicolumn{1}{c|}{$++$} & $--$ & \multicolumn{1}{c|}{} &  & $\downarrow$ &  \\ \hline
3 & 0 & \multicolumn{1}{c|}{$++$} & \multicolumn{1}{c|}{$--$} & $++$ & \multicolumn{1}{c|}{$--$} & $--$ & $\uparrow$ & $\uparrow$ \\ \hline
4 & 1 & \multicolumn{1}{c|}{$--$} & \multicolumn{1}{c|}{$++$} & $--$ & \multicolumn{1}{c|}{$++$} & $++$ & $\downarrow$ & $\downarrow$ \\ \hline
\end{tabular}
\end{adjustbox}
\caption{Four types of noise events in $b_i$ affect surrounding bits. Here, $j \in \{1, 2\}$.}
\vspace{-3mm}
  \label{tab:4type}
\end{table}


For short noise events, large noises at time $i$ cause an increase (type 1) or decrease (type 2) in the sum of one-time-step parity bit $y_{i+1,1} + y_{i+1,2}$ (second-order for $b_i$, third-order for $b_{i-1}$). This results in over-correction when decoding the previous bit $b_{i-1}$ with minimal noises, as shown in Fig. \ref{fig:past}. 
\begin{figure}[ht]
\vspace{-4mm}
    \centering
    \includegraphics[width=0.9\columnwidth]{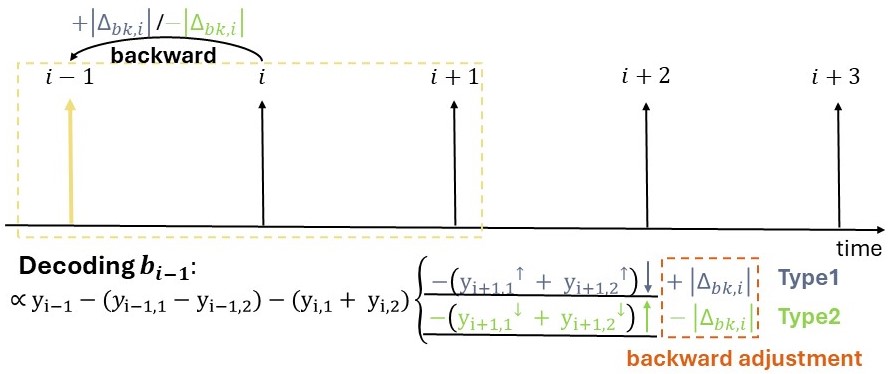}
    \caption{Effect of problematic bit $b_i$ on the previous bit}
    \vspace{-3mm}
    \label{fig:past}
\end{figure}


{For long noise events (type 3 or type 4), changes in the two-time-step parity bit sum $(c_{i+2,1} + c_{i+2,2})$ aid in correctly decoding $b_i$ but can lead to unnecessary correction for $b_{i+1}$, as illustrated in Fig. \ref{fig:future}.} Although entanglement alleviates this issue to some extent, it is insufficient. We need to address these problems with more care, including forward and backward adjustments, which will be discussed later.

 \begin{figure}[ht]
    \centering
    \includegraphics[width=0.9\columnwidth]{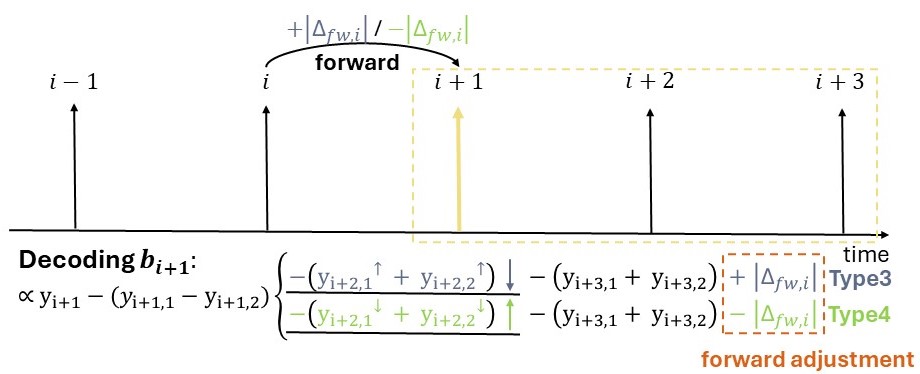}
    \caption{Effect of problematic bit $b_i$ on the subsequent bit}
    \vspace{-3.5mm}
    \label{fig:future}
\end{figure}





{\bf Two-stage decoding for interpretable enc 3:} To solve the over-correction issue, we present a two-stage, interpretable decoder inspired by the two-layered bi-directional GRU in the original Deepcode structure. This design cleverly separates the ``\textcolor{cyan}{second-order}'' or ``\textcolor{orange}{third-order}'' components for different bits, identifying the problematic bit and signaling surrounding bits.

\subsubsection{First-stage}
In the first stage of decoding, the decoder performs first-order error correction at each time step $i$:
\begin{equation}
    g_{i,p} = \tanh\left(\textcolor{red}{\alpha_{p,1}}y_i - \textcolor{red}{\alpha_{p,2}}(y_{i, 1}- y_{i, 2})\right), \quad p \in \{1, 2, 3\}
\end{equation}
where $\alpha_{p,1}$ and $\alpha_{p,2}$ are learned parameters. The decoding process removes phase 1 noise (estimated by subtracting the current parity bits) from the uncoded noisy $y_{i}$, yielding a preliminary decoding result for further refinement in the second stage. This typically results in correct decoding unless the phase 2 noise is excessive.

\subsubsection{Second-stage}

In the second stage, the decoder performs a bi-directional decoding process, adding higher-order error correction terms to decode the message bit. Similar to single-stage decoding, the idea is that the sum of one-time-step parity bits ($y_{i+1,1} + y_{i+1,2}$) corrects large phase 2 noises with different signs at time $i$, while the sum of two-time-step parity bits ($y_{i+2,1} + y_{i+2,2}$) prevents perturbations from same-sign phase 2 noises at $i+1$. Additionally, we consider the sum of three-time-step parity bits ($y_{i+3,1} + y_{i+3,2}$) to account for potential entanglement. As shown in Fig. \ref{fig:future}, the forward pass adjusts the decoding of subsequent bits, which incorporates information from time $i+3$. However, as shown in Table \ref{table:decoder compare}, extending the single-stage decoding from dec 3 to dec 4 yields negligible performance improvement. This highlights the necessity of the forward and backward pass to adjust the over-correction. We denote the forward direction state as ``$fw$'' and the backward direction state as ``$bk$'' using subscripts $i$ for each time step and $q$ for different states. In the following, we do not require the positivity of the learned parameters.


In the forward direction, we assess whether prolonged noise events induce type 3 or type 4 error corrections. This assessment involves comparing forward decoding outcomes by adjusting the sum of two-time-step parity bits: either adding or subtracting this sum $(y_{i+2,1}+y_{i+2,2})$, and evaluating the resulting differences to determine their effects. 
\begin{align}
fw_{i,q} = & \tanh \Bigl( \sum_{p=1}^{3}\textcolor{red}{\beta_{q, p}}g_{i,p} - \textcolor{red}{\beta_{q, 4}} (y_{i+1,1}+y_{i+1,2}) \Bigr. \nonumber  \\
& \Bigl.\textcolor{red}{-} \textcolor{red}{\beta_{q, 5}}(y_{i+2,1}+y_{i+2,2}) - \textcolor{red}{\beta_{q, 6} }(y_{i+3,1}+y_{i+3,2}) \Bigr) \label{eq:forwardstate}
\end{align}

We compare the decoding results using the formula:
\begin{equation}
    \Delta_{fw,i,q} = \textcolor{red}{\gamma_{q,1}} fw_{i,1} + \textcolor{red}{\gamma_{q,2}} fw_{i,2} + \textcolor{red}{\gamma_{q,3}} fw_{i,3}
\end{equation}
where $q \in \{1, 2, 3\}$. If no type 3 or type 4 errors occur, adding or subtracting the sum of two-time-step parity bits (third-order error correction of $b_i$) will yield nearly identical decoding results for $b_i$. Therefore, $\Delta_{fw,i,q}$, which denotes the difference in the outcome of forward states, will be negligible, approximately $0$. If $b_i$ exhibits type 3 (or type 4) noise events, the decoding results will differ significantly with or without third-order error correction. Therefore, $\Delta_{fw,i} \gg 0 $ (or $\Delta_{fw,i} \ll 0 $) will have an extreme positive (or negative) outlier, which will be passed to the subsequent bit to prevent over-correction, as depicted in Fig. \ref{fig:future}. This leads to the formation of new forward ``beliefs'' regarding the subsequent bits:
\begin{equation}
    fw_{i+1, q}^{'} = \tanh\left(\tanh^{-1}(fw_{i+1,q}) + \textcolor{teal}{\Delta_{fw,i,q}}\right) \label{eq:newforward}
\end{equation}




In the backward direction, backward decoding results are similarly compared by adjusting the sum of one-time-step parity bits ($y_{i+1,1} + y_{i+1,2}$) to determine whether to signal the previous bit:
\begin{align}
\label{eq:backwardstate}
bk_{i,q} = & \tanh \Bigl( \sum_{p=1}^{3}\textcolor{red}{\beta_{q, p}}g_{i,p} - \textcolor{red}{\beta_{q, 4}} (y_{i+1,1}+y_{i+1,2}) \Bigr. \nonumber \\
& \Bigl.\textcolor{red}{-} \textcolor{red}{\beta_{q, 5}}(y_{i+2,1}+y_{i+2,2}) - \textcolor{red}{\beta_{q, 6} }(y_{i+3,1}+y_{i+3,2}) \Bigr)  \\
\Delta_{bk,i,q} =&  \textcolor{red}{\gamma_{q,1}} bk_{i,4} + \textcolor{red}{\gamma_{q,2}} bk_{i,5} + \textcolor{red}{\gamma_{q,3}} bk_{i,6}
\end{align}
where $q \in \{4, 5, 6\}$. If $b_i$ exhibits no type 1 or type 2 errors, adjusting (adding or subtracting) the sum of one-time-step parity bits (second-order correction) has minimal effect ($\Delta_{bk,i,q}$ is small). However, with type 1 or type 2 errors, this adjustment significantly alters the decoding results, resulting in a large positive or negative $\Delta_{bk,i}$ that will be passed to the previous bit for further refinement (see Fig. \ref{fig:past}). The updated backward ``beliefs'' on the previous bit are expressed as:
\begin{equation}
    bk_{i-1, q}^{'} = \tanh\left(\tanh^{-1}(bk_{i-1,q}) + \textcolor{teal}{\Delta_{bk,i,q}} \right)
\end{equation}

{\bf Interestingly, the decoding process gathers information from surrounding bits (through forward and backward passes) to mitigate over-correction.} Consequently, we compute a linear combination of updated forward and backward beliefs to obtain a soft output, which is subsequently thresholded to derive our estimated bits:
\begin{align}
   \begin{adjustbox}{max width = \columnwidth}
$
\hat{b}_{i}  =  \begin{cases}0 & \text{if } D_i < 0.5, \\ 1 & \text{if } D_i \geq 0.5. \end{cases} \quad D_i  = \sigma\left(\sum_{q=1}^{3}\textcolor{red}{r_q}fw_{i,q}^{'} + \sum_{q=4}^{6}\textcolor{red}{r_q}bk_{i,q}^{'}\right)
$
\end{adjustbox}     
\end{align}
where $r_q$ represents learned coefficients.

We evaluate various decoder architectures with a fixed encoder, specifically interpretable enc 3, as shown in Table \ref{table:decoder compare}. The results show that the two-stage dec 4 mitigates over-correction and achieves comparable BER performance to the deep learned decoder with 5 hidden states, which is significantly better than the single-stage decoder.
\begin{table}[ht]
\centering
\begin{adjustbox}{max width = \columnwidth}
\begin{tabular}{|l|l|l|}
\hline
Encoder & Decoder & BER  $\text{SNR}_f$ $0\text{dB}$    \\ \hline
 interpret. enc 3  & interpret. dec 2 single-stage    &  $7.468e-05$   \\ \hline
 interpret. enc 3  & interpret. dec 3 single-stage & $3.292e-05$ \\ \hline
 interpret. enc 3 & interpret. dec 4 single-stage  &  $2.906e-05  $ \\ \hline
 interpret. enc 3 & interpret. dec 4 two-stage &  $8.587e-06 $\\ \hline
\end{tabular}
\end{adjustbox}
\caption{\label{table:decoder compare}BER performance of different decoders (noiseless feedback).}
\vspace{-3mm}
\end{table}

In conclusion, the (enc 3, dec 4) two-stage interpretable model establishes longer dependencies between codewords and incorporates information from surrounding bits through forward and backward passes.

\subsection{Noisy Feedback}
In this subsection, we extend our interpretable model to the noisy feedback case. We begin by investigating the pruning effect on the learned parameters of the encoder RNN in Deepcode, which has 7 hidden states, across various feedback SNRs. The RNN is defined as $\mathbf{h}_{i} = \tanh(\textcolor{red}{\mathbf{W_{hp}}}\mathbf{P}_{i} + \textcolor{red}{\mathbf{W_{hh}}}\mathbf{h}_{i-1}+ \textcolor{red}{\mathbf{b}})$, where $\mathbf{W_{hp}}, \mathbf{W_{hh}}$ are the learned parameter matrices and $\mathbf{b}$ is the bias. The input $\mathbf{P}_{i}$ is of size four, containing message bit $b_i$, phase 1 noises $n_i+ \tilde{n}_i$ and delayed phase 2 noises $n_{i-1,1} + \tilde{n}_{i-1,1}$ and $n_{i-1,2} + \tilde{n}_{i-1,2}$. Pruning involves sequentially setting learned parameters with smaller absolute values to 0.

As shown in Fig. \ref{fig:pruning_noisy}, for all feedback SNRs, the maximum acceptable pruning level for the input-to-hidden parameter matrix $\mathbf{W_{hp}}$ is 50\%; exceeding this limit significantly worsens BER performance. However, as the feedback becomes noisier, pruning the hidden-to-hidden parameter matrix $\mathbf{W_{hh}}$ has less impact on BER performance. This suggests that when feedback is less reliable, the encoder incorporates it to a lesser degree in the parity bits.

\begin{figure}[ht]
\vspace{-4mm}
    \centering
    \includegraphics[width=0.9\columnwidth]{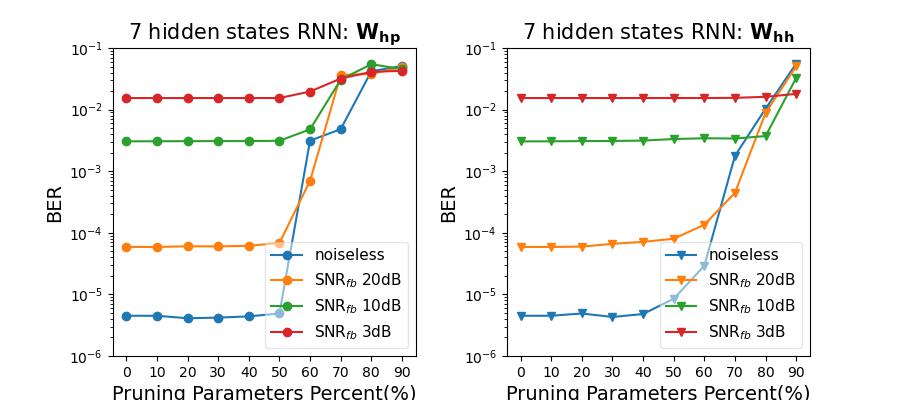}
    \caption{BER Performance of Pruning in Deepcode
with 7 hidden states ($\text{SNR}_f = 0\text{dB}$, noisy feedback)}
    \vspace{-2mm}
    \label{fig:pruning_noisy}
\end{figure}

Through pruning and outlier analysis, we observe that the difference in the interpretable model between noiseless and noisy feedback is in the first-order error correction component. Unlike noiseless feedback with fixed knee points at the origin, noisy feedback exhibits varying knee points based on the message bit $b_i$, as noted in \cite{US-2024-ISIT-IEEE}. Therefore, in the case of noisy feedback, the first-order error correction is:
\begin{align}
 &\text{If } b_{i}=0, (n_{i}+\tilde{n}_{i} + \textcolor{red}{\lambda_1})\mathbb{I}(-(2b_i-1)(n_{i}+\tilde{n}_{i} + \textcolor{red}{\lambda_1}))\\
 & \text{If } b_{i}=1, (n_{i}+\tilde{n}_{i} - \textcolor{red}{\lambda_2})\mathbb{I}(-(2b_i-1)(n_{i}+\tilde{n}_{i} - \textcolor{red}{\lambda_2}))
 \end{align}
 where $\lambda_1$ and $\lambda_2$ are learned parameters. {As the feedback becomes noisier, outliers in second-order and third-order error correction will increase due to more frequent noise events.}

\section{Empirical Results}\label{sec: Empirical Results}
In this section, we discuss the training and results of the interpretable models\footnote{Our code: \url{https://github.com/zyy-cc/Deepcode-Interpretable-higher}} with noiseless or noisy feedback. During training, we optimize all model parameters by minimizing the BCE across different forward or feedback SNRs. Our implementation is in PyTorch, while the original Deepcode with 50 hidden states is implemented in TensorFlow. Compared to the original Deepcode with over 65,000 parameters, the (enc 3, dec 4) two-stage model has 90 parameters with fixed knee points (92 with varying knee points), including 12 for power allocation. This makes the interpretable model significantly more computationally efficient. 

The BER performance with noiseless feedback is illustrated in Fig. \ref{fig:7berperformance}. The results indicate that the (enc 3, dec 4) two-stage interpretable model outperforms the (enc 2, dec 2) single-stage model at low forward SNRs. However, at $\text{SNR}_f$ $-1$dB, it remains inefficient in handling significant noise levels compared to Deepcode with 50 hidden states. Conversely, at high $\text{SNR}_f$ $2$dB, the (enc 3, dec 4) two-stage model is inferior to the (enc 2, dec 2) single-stage model, {which we suspect is due to either a) that training a model with more parameters becomes challenging at high SNR when there are fewer errors, b) long-dependency is not needed at high SNR, {as lower-order error correction adequately handles noise for correct decoding.}}


\begin{figure}[ht]
 \vspace{-5mm}
    \centering
    \includegraphics[width=0.9\columnwidth]{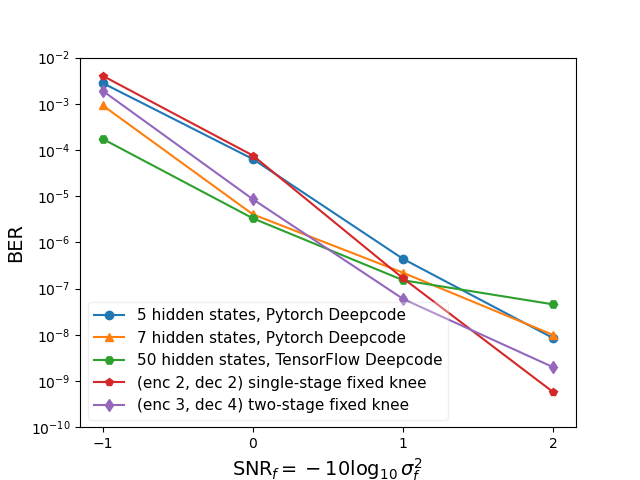}
    \caption{BER performance (noiseless feedback) as a function of the forward SNR.}
    \vspace{-2mm}
    \label{fig:7berperformance}
\end{figure}

Fig. \ref{fig:noisyperformance} shows the BER performance for AWGN channel with noisy feedback. At high $\text{SNR}_{fb}$, the (enc 3, dec 4) two-stage interpretable model outperforms the (enc 2, dec 2) single-stage model. As the $\text{SNR}_{fb}$ decreases, all models converge to similar BER performance due to less reliable feedback. {Interpretable models with varying knee points slightly outperform those with fixed knee points due to their increased flexibility in transmitting phase 1 noise based on the condition of the feedback channel.}

\begin{figure}[ht]
\vspace{-4mm}
    \centering
    \includegraphics[width=0.9\columnwidth]{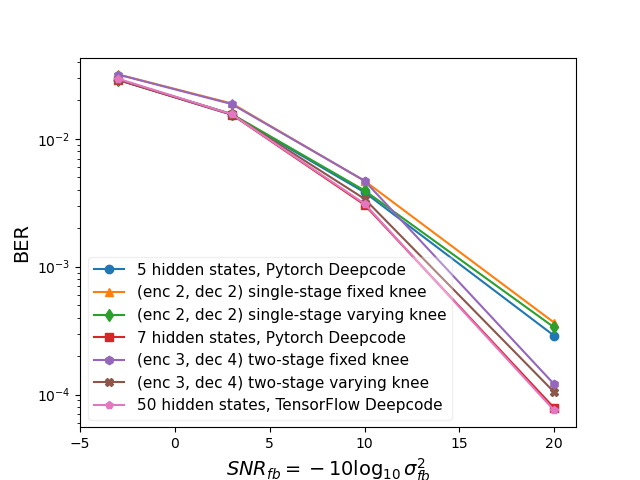}
    \caption{BER performance ($\text{SNR}_f = 0$dB, noisy feedback) as a function of the feedback SNR}
    \vspace{-2mm}
    \label{fig:noisyperformance}
\end{figure}

\section{Conclusions}
We present an (enc 3, dec 4) two-stage interpretable model, which improves BER performance over our previous (enc 2, dec 2) single-stage model. The encoder uses higher-order error correction for longer noise events, while the decoder employs bidirectional passes to reduce over-correction.  {One key takeaway from our interpretable model of Deepcode is that it cleverly performs error correction within the structure enforced by Deepcode's architecture (two-phases, rate $1/3$, feedback encoder RNN and bidirectional GRU decoder structures). The decoder essentially works by trying to subtract off the noise of the first uncoded transmission phase; the encoder codes information in order to help it do so.  The decoder adds or subtracts parity bit pairs, but in doing so additional noise terms may be added, and those are corrected by subsequent parity bits. The challenge lies in cleverly combining the decoding and error correction of bit $b_i$ with those of bit $b_{i-1}$ and bit $b_{i+1}$ (and so on) for longer and longer sequences of noises and memory indicated by the encoder -- {this is enabled through a two-stage decoder where forward and backward adjustments identify error types and prevent over-correction}. In short, more hidden RNN states appear to equate with longer memory lengths and longer sequences of error correction, {or higher order error correction}.} 


\bibliographystyle{IEEEtran}
\bibliography{reference}

\newpage
\appendices

\end{document}